\documentclass[fleqn,usenatbib]{mnras}

\usepackage{newtxtext,newtxmath}

\usepackage[T1]{fontenc}

\DeclareRobustCommand{\VAN}[3]{#2}
\let\VANthebibliography\thebibliography
\def\thebibliography{\DeclareRobustCommand{\VAN}[3]{##3}\VANthebibliography}

\usepackage{graphicx}	
\usepackage{amsmath}	

\title[]{Universal relations to measure neutron star properties from targeted r-mode searches}

\author[S. Ghosh]{
Suprovo Ghosh$^{1}$\thanks{E-mail: suprovo@iucaa.in}
\\
$^{1}$Inter-University Centre for Astronomy and Astrophysics,
Pune University Campus,
Pune 411007, India
}

\date{Accepted XXX. Received YYY; in original form ZZZ}

\pubyear{2015}

\begin{document}
\label{firstpage}
\pagerange{\pageref{firstpage}--\pageref{lastpage}}
\maketitle

\begin{abstract}
R-mode oscillations of rotating neutron stars(NS) are promising candidates for continuous gravitational wave (GW) observations. In our recent work, we derived universal relations of the NS  parameters, compactness and dimensionless tidal deformability with the r-mode frequency. In this work, we investigate how these universal relations can be used to infer various NS  intrinsic parameters following a successful detection of the r-modes. In particular, we show that for targeted r-mode searches, these universal relations along with the ``I-Love-Q'' relation can be used to estimate both the moment of inertia and the distance of the NS thus breaking the degeneracy of distance measurement for continuous gravitational wave(CGW) observations. We also discuss that with a prior knowledge of the distance of the NS from electromagnetic observations, these universal relations can also be used to constrain the dense matter equation of state (EOS) inside NS. We quantify the accuracy to which such measurements can be done using the Fisher information matrix for a broad range of possible, unknown parameters, for both the a-LIGO and Einstein Telescope (ET) sensitivities. 
\end{abstract}

\begin{keywords}
stars: oscillations --  pulsars: general -- equation of state -- gravitational waves
\end{keywords}



\section{Introduction}
The first detection of gravitational waves from a binary black hole(BBH) merger event GW150914~\citep{LIGOGW1st} has opened up a new rapidly growing field in modern astrophysics. Since the first detection of a BBH merger in 2015, till now we have made almost 100 confirmed detections~\citep{GWcatalouge} of gravitational waves from binary systems including binary neutron star (BNS)~\citep{Abbott2017} and neutron star-black hole (BH-NS)~\citep{Abbott_2021} systems using the LIGO–Virgo–Kagra(LVK) global network of GW detectors~\citep{AdvLIGO2015,Abbott2016,AdvVIRGO2014,Kagra}. The BNS merger event GW170817~\citep{Abbott2017,Abbott2018,Abbott2019} was also observed through out the electromagnetic spectrum opening the modern era of multi-messenger astronomy~\citep{MultiGW170817}. \\

Although till now, we have only observed GW emission from binary sources, with increasing sensitivity of the current detectors and also the upcoming third-generation detectors like Einstein telescope~\citep{ET1,ET2} or Cosmic Explorer~\citep{CE}, we will also possibly detect weaker sources of gravitational waves such as continuous gravitational waves. CGW emission from spinning isolated NSs are characterised by their long-lasting and almost-monochromatic nature. Deformations or ``mountains" on the surface of NS supported by elastic and/or magnetic strain, rigid rotation of a triaxial star or fluid oscillations in a rotating NS are the main sources of CGW emission from NS (See~\cite{Riles_2023} for a recent review). \\

R-mode is a toroidal mode of fluid oscillation for which the restoring force is the Coriolis force~\citep{rmode1977,Andersson1998,Friedman:1997uh,Andersson2003}. For any rotating star, Chandrasekhar–Friedman–Schutz~\citep{CFS1,CFS2} mechanism drives the r-mode unstable leading to GW emission. This instability can explain the spin-down of hot and young NSs~\citep{Andersson1999,Lindblom1998,Alford2014},as well as old, accreting NSs in low-mass X-ray binaries~\citep{Bildsten:1998ey,Andersson:1998qs,Wynn2011}. Because of the astrophysical significance, CGW emission from r-modes has been searched in the LVK data for Crab pulsar~\citep{Rajbhandari2021} and PSR J0537-6910~\cite{Fesik2020a,Fesik2020b,LIGO_rmode}. No CGWs are still detected in these searches, but upper limits on strain amplitude were obtained. \\

GW emission from binary systems allows to calculate the luminosity distance of the system from the measured GW parameters alone~\citep{Schutz1986}, that's why these systems are called ``standard siren". Recently,~\cite{Sieniawska2021} showed that CGW emission from NS can't be used as standard sirens since the distance estimation is always degenerate by one of the unknown physical parameters : moment of inertia (MoI) or the r-mode amplitude, denoted by $\alpha$ (ellipticity, $\epsilon$ in case of ``mountains"). In the specific case of r-modes from barotropic, slowly rotating NS, the frequency of the emitted GW actually depends on the structure of the NS~\citep{Lindblom:1999yk,Lockitch2001,Lockitch2003}. In our recent paper~\citep{Ghosh2023}, we improved the universal relation of the r-mode frequency with the NS compactness considering the Newtonian limits of the r-mode frequency and recent multi-messenger constraints on the NS EOS~\citep{Traversi2020,Dietrich2020,Pang_2021,Ghosh_2022,Ghosh2022} and also derived universal relation between r-mode frequency and dimensionless tidal deformability of the NS.  The current most-sensitive searches for r-mode using the LVK detectors are targeted searches for which we know the rotation frequency of the pulsars and the search is within a band of few Hz~\citep{Caride2019}. In this paper, we show that for such targeted r-mode searches, from the measured frequency we can determine the MoI of the system for a given EOS and then calculate the distance thus breaking the degeneracy between MoI and distance. We further analyse the accuracy to which such measurements can be made with a-LIGO detector~\citep{ALIGO,a-LIGO} and third-generation detectors like the Einstein Telescope. In particular, we show how the uncertainty in the inclination angle($\iota$) will affect the distance measurement of the pulsar.  Although we do not include simulations for the third generation detector Cosmic Explorer~\citep{CE,CE2} but the results are expected to be qualitatively similar to the Einstein Telescope. We also discuss that if we assume the knowledge of the distance from prior EM observations, we can use the universal relations to measure the mass and MoI of the NS that can be used to constrain the EOS.\\

The structure of the article is as follows: in Sec.~\ref{sec:formalism}, we describe the formalism of CGW from r-modes and the universal relations for the r-mode frequency. We also estimate errors of relevant signal parameters using the Fisher matrix. The details of the measured NS parameters and corresponding errors are discussed in Sec.~\ref{sec:results}. In Sec.~\ref{sec:assumptions}, we discuss the main assumptions considered in this study and their validity with current observation scenarios. In Sec.~\ref{sec:discussion} we discuss the main implications and future aspects of this work.

\section{Method and Formalism}
\label{sec:formalism}

\subsection{Continuous gravitational waves from r-modes}
\label{sec:CGW}
 The strain amplitude for r-mode oscillations is parameterised by the dimensionless quantity $\alpha$~\citep{Owen1998} and the corresponding CGW strain is given by~\citep{Owen2010} 
\begin{equation}\label{eq:ramp}
    h_0 = \sqrt{\frac{512\pi^7}{5}}\frac{G}{c^5}(\alpha MR^3\Tilde{J})\frac{1}{d}f_{GW}^3
\end{equation}
where $M$, $R$ are the mass and radius of the star respectively; $f_{GW}$ is the frequency of the gravitational wave ; $\alpha$ is the r-mode saturation amplitude and $\Tilde{J}$ is a dimensionless parameter that depends on the EOS of the NS~\citep{Owen1998},
\begin{equation}\label{eq:j}
    \Tilde{J} = \frac{1}{MR^4}\int_0^R \rho r^6 dr
\end{equation}
where $\rho$ is the density inside the NS. \\ 
Assuming the star is losing all of its rotational energy via r-mode emission and the constant r-mode saturation amplitude $\alpha$, the frequency derivative can be written as~\citep{Riles_2023}
\begin{equation}\label{eq:spindown}
   \dot{f}_{GW} = -\frac{4096\pi^7}{25}\frac{G}{c^7}\frac{M^2R^6\Tilde{J}^2}{I}\alpha^2f_{GW}^7 
\end{equation}
where $I$ is the MoI of the NS. Eliminating the highly uncertain parameter $\alpha MR^3\Tilde{J}$ from Eqn.~\eqref{eq:ramp} and Eqn.~\eqref{eq:spindown}, we have
\begin{equation}\label{eq:moid}
    \frac{\sqrt{I}}{d} = h_0\sqrt{\frac{f_{GW}}{\dot{f}_{GW}}} \sqrt{\frac{8c^3}{45G}} .
\end{equation}
This means that, unlike the binary inspirals~\citep{Schutz1986}, we cannot directly solve for distance ($d$) from the gravitational wave measurement alone and it will always be degenerate with the MoI ($I$)~\citep{Sieniawska2021}. 

\subsection{Universal relations}
\label{sec:UR}
For slowly and uniformly rotating barotropic stars, the r-mode frequency is proportional to the rotational frequency of the NS~\citep{idrisy2015,Ghosh2023}, 
\begin{equation}\label{eq:freq}
    f_{GW} = |2-\kappa|f_{rot}
\end{equation}
Where $f_{rot}$ is the rotational frequency of the star and $\kappa$ is a dimensionless parameter. ~\cite{idrisy2015} showed that $\kappa$ has a universal relation  with the neutron star compactness($C$) which we improved considering a chosen set of 15 tabulated EOSs that are consistent with the recent multi-messenger observations of NS and the Newtonian limits of the r-mode frequency~\citep{Ghosh2023} 
\begin{equation}\label{univ_real}
    \kappa = 0.667 - 0.478C - 1.11C^2.
\end{equation}
We also showed that $\kappa$ has a universal relation with the dimensionless tidal deformability($\Lambda$)~\citep{Ghosh2023}
\begin{equation}\label{eq:univ_lamb}
    \kappa = 0.3612 + 0.0407 \> log(\Lambda) - 0.0015 \> log^2(\Lambda) .
\end{equation}
For slowly rotating stars, there are also universal relations between the NS MoI, the Love numbers and the quadrupole moment called the \textit{I-Love-Q} relations~\citep{Yagi2013}. The universal relation between the normalised MoI($\Bar{I} = I/m^3$) and tidal deformability is given by~\citep{yagiYunes2013}
\begin{equation}\label{eq:univ_IloveQ}
\begin{split}
    ln(\Bar{I}) = 1.47 + 0.0817 \> ln(\Lambda) + 0.0179 \> ln(\Lambda)^2  + 2.87\times 10^{-4} \> ln(\Lambda)^3 \\ - 3.64\times 10^{-5} \> ln(\Lambda)^4
\end{split} 
\end{equation}

\subsection{Signal model and error estimation using Fisher matrix}
\label{sec:Fisher}
The strain produced in a detector by the continuous gravitational wave signal from the r-mode oscillations in neutron star can be represented in the form of~\citep{JK1998}
\begin{equation}\label{eq:strain}
    h(t) = \sum_{i=1}^{4}\mathcal{A}^ih_i(t;\lambda)
\end{equation}
in terms of four signal-amplitudes $\mathcal{A}$ independent of the detector and the detector-dependent basis $h(t;\lambda)$. The signal amplitudes $\mathcal{A}^i$ can be expressed in terms of the two polarization amplitudes $A_{+},A_{\times}$, the initial phase $\Psi_0$ and the polarization angle $\psi$. The additional parameters, represented by $\lambda$ in equation~\eqref{eq:strain} which modify the phase of the signal, include the star’s sky location, detector position and, if the star is in a binary system,its orbital parameters.  The polarisation amplitudes $A_{+}$ and $A_{\times}$ can be written in terms of the characteristic amplitudes $h_0$ and the inclination angle $\iota$ between the neutron star’s rotation axis to the line of sight~\citep{JK1998}
\begin{equation}\label{eq:incl}
    A_{+} = \frac{1}{2}h_0(1 + \cos^2{\iota}) \text{\hspace*{3mm}} A_{\times} = h_0\cos{\iota} .
\end{equation}
\\
To express the phase of the continuous gravitational wave signal, we assumed that in the rest frame  of  the  neutron  star  the  gravitational wave frequency  can  be  expanded  in  a  Taylor  series. For the case of targeted searches (which is also we are considering in this work for most sensitive r-mode searches), the sky location of the source is known and the phase can be expressed as a polynomial function of the initial phase ($\Psi_0$), frequency and the higher order derivatives~\citep{JK1999},
\begin{equation}\label{eq:phase}
     \Psi = \Psi_0 + 2\pi\left[ ft + \frac{1}{2}\Dot{f}t^2 + \frac{1}{6}\Ddot{f}t^3\right]
\end{equation}
where t is an arbitrary time and $\Psi_0$ is the initial phase which is set to be zero. The parameter space metric over the phase parameter set $\mathcal{P} = \left(f,\Dot{f},\Ddot{f}\right)$ can be written as 
\begin{equation}\label{eq:parameter}
    g_{ij} = \Biggl \langle \frac{\partial \Psi}{\partial f^{(i)}}\frac{\partial \Psi}{\partial f^{(j)}} \Biggr \rangle -\Biggl \langle\frac{\partial \Psi}{\partial f^{(i)}}\Biggr \rangle \Biggl \langle \frac{\partial \Psi}{\partial f^{(j)}} \Biggr \rangle
\end{equation}
where $ \left( f^{(0)},f^{(1)},f^{(2)}\right) = \left(f,\Dot{f},\Ddot{f}\right)$. The Fisher Co-variance matrix is given by the inverse of the above matrix 
\begin{equation}\label{eq:fisher}
    \Gamma = \frac{g^{-1}}{\rho^2}
\end{equation}
where $\rho^2$ is the signal-to-noise ratio(SNR) assuming optimal match between the true signal and the best-fit template~\citep{Moore2015}. The SNR is calculated using the formula
\begin{equation}\label{eq:snr}
    \rho^2 = \int_0^{\infty} \frac{4|\Tilde{h}(f)|^2}{S_n(f)} df
\end{equation}
where $\Tilde{h}(f)$ is a Fourier transform of the gravitational wave signal and $S_n(f)$ is the amplitude spectral density which determines the sensitivity of the detector. For observation times of one year or more, we can get an expression for SNR averaged over the sky location and polarisation angle $\psi$~\citep{JK1998} 
\begin{equation}\label{eq:final_snr_pol}
    \rho^2 = \frac{1}{40}\frac{h_0^2T}{S_n(f)}(1+6\cos^2{\iota} + \cos^4{\iota}),
\end{equation}
and if we further average over inclination angle $\cos{\iota} \in [-1,1]$, then we get~\citep{Prix2011}, 
\begin{equation}\label{eq:final_snr}
    \rho^2 = \frac{4}{25}\frac{h_0^2T}{S_n(f)}.
\end{equation}
The error propagation for any physical quantity that depends on our parameter set $\mathcal{P} = \left(h_0, f,\Dot{f},\Ddot{f}\right)$ can be written as 
\begin{equation}\label{eq:error}
   \sigma (A)^2 = \sum_{x,y \in \mathcal{P} } \left(\frac{\partial A}{\partial x}\right) \left(\frac{\partial A}{\partial y}\right) \Gamma_{xy}
\end{equation}
where $\Gamma_{xy}$ denotes the co-variance between the variables $x,y$. The error for the frequency and spin down parameters can be obtained by evaluating Eqn.~\ref{eq:parameter} with the expression for the phase $\Psi$ given in Eqn.~\eqref{eq:phase}. For the amplitude parameters, only $h_0$ parameter is of importance to infer the NS interior properties. For targeted and year-long searches, we can consider the error in $h_0$ averaged over sky position and polarisation angle $\psi$ but dependent of the inclination angle $\iota$,
\begin{equation}\label{eq:error_h}
    \frac{\sigma(h_0)}{h_0} = \frac{2a}{5\rho}\frac{\sqrt{b+\cos^2{i}}}{1-\cos^2{i}}
\end{equation}
where $a \approx 4.08$ and $b\approx 2.59$~\citep{Lu_2023,Prix2011}. Due to the singularity in the coordinate transformation between the four amplitude parameters $A_i$ and $\{h_0,\iota,\psi,\Psi_0\}$, there is  a divergence in the error in Eqn.~\eqref{eq:error_h} for $\cos{\iota} = \pm 1$~\citep{Prix2011}. Using the formula~\eqref{eq:error}, we get the error estimates for the quantity $\frac{d}{\sqrt{I}}$ from the Eqn.~\eqref{eq:moid}~\citep{Sieniawska2021}. 
\begin{equation}\label{eq:errorfrac}
\begin{split}
    \sigma\left(\frac{d}{\sqrt{I}}\right)^2 &= \frac{45G}{8c^3}\frac{1}{(\pi\rho h_0)^2}\left[\frac{75}{T}\frac{\Dot{f}}{f^3} + \frac{1620}{T^4f\Dot{f}} + \frac{675}{T^3f^2} + \right. \\ & \left. \pi^2\frac{\Dot{f}}{f}\frac{2a}{5}\frac{\sqrt{b+\cos^2{i}}}{1-\cos^2{i}}\right]
\end{split}    
\end{equation}
The parameter $\kappa$ is calculated from the measured gravitational wave frequency $(f_{GW})$ using the Eqn.~\eqref{eq:freq}. The error in measuring the parameter $\kappa$ is estimated by 
\begin{equation}\label{eq:error_k}
\sigma (\kappa)^2 = \left(\frac{f_{GW}}{f_{rot}}\right)^2\left[ \left(\frac{\sigma (f_{GW})}{f_{GW}}\right)^2 + \left(\frac{\sigma (f_{rot})}{f_{rot}}\right)^2 \right]
\end{equation}
From the universal relations~\eqref{eq:univ_lamb}~\eqref{eq:univ_IloveQ}, we can estimate the error in the normalised MoI to be 
\begin{equation}\label{eq:error_I}
\begin{split}
    \frac{\sigma (\Bar{I})^2}{\Bar{I}^2} = \left[ (0.0817+0.0358 ln\Lambda+0.0008ln\Lambda^2 - 0.0002ln\Lambda^3) \right] \\
    \times  \sigma (\kappa)^2\frac{1}{0.0049 - 0.008\kappa}
\end{split} 
\end{equation}
which can be converted into $\sigma (I)/I$ for a given EOS. Now, the error is measuring the distance can be simply calculated from~\eqref{eq:error_I} and ~\eqref{eq:errorfrac}
\begin{equation}\label{eq:error_d}
    \left(\frac{\sigma (d)}{d}\right)^2 = \left(\frac{\sigma (d/\sqrt{I})}{d/\sqrt{I}}\right)^2 + \frac{1}{4}\left(\frac{\sigma (I)}{I}\right)^2.
\end{equation}
\begin{figure}
\centering
\includegraphics[width=.5\textwidth]{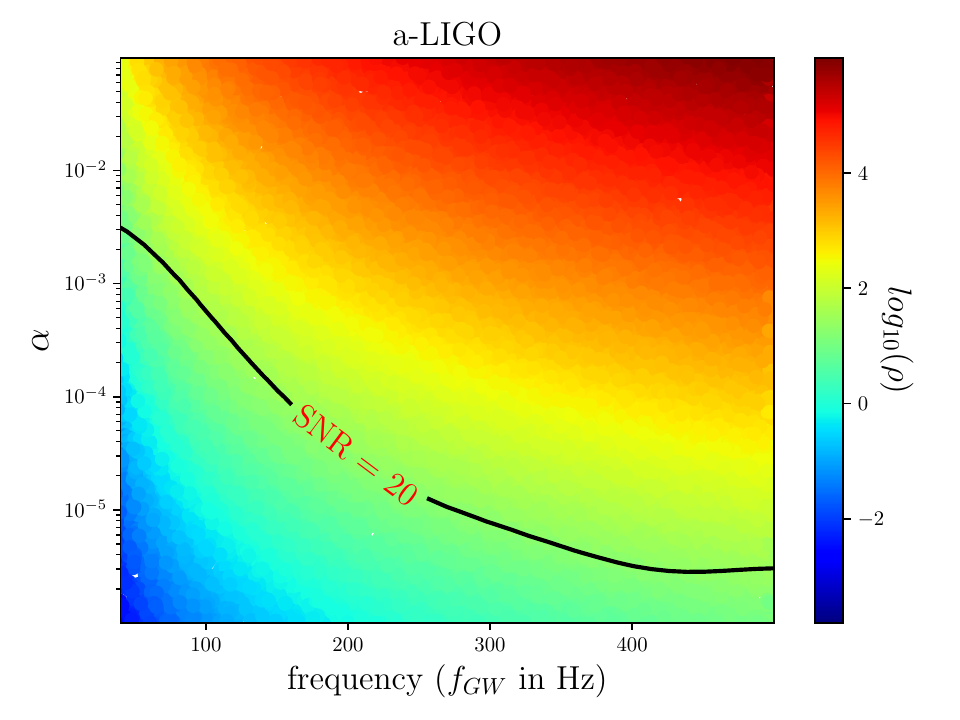}\hfill
\includegraphics[width=.5\textwidth]{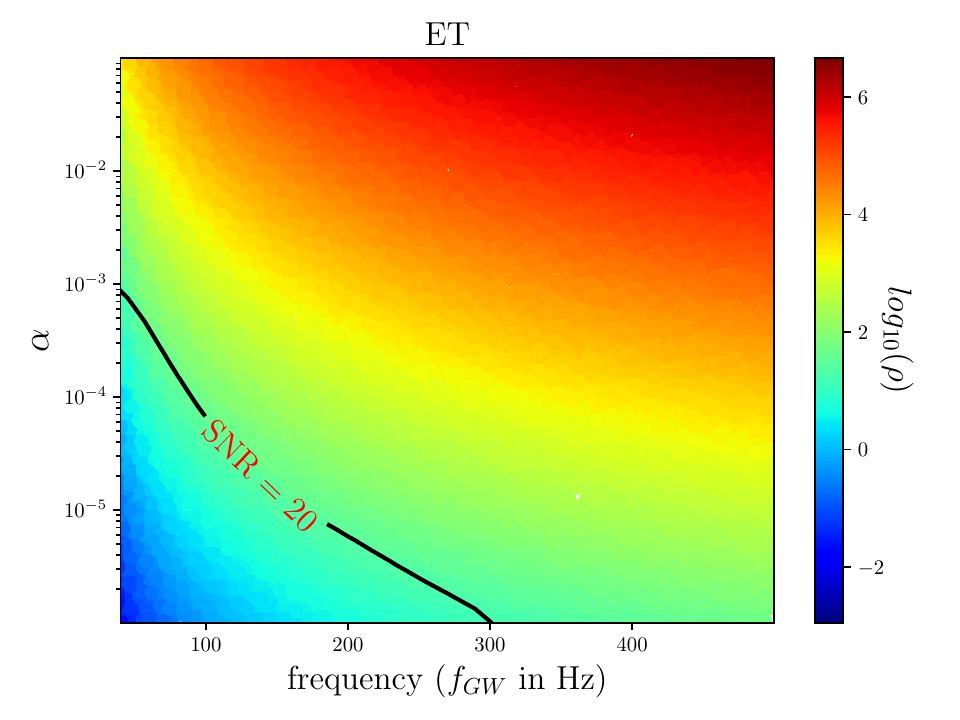}
\caption{Expected SNR($\rho$) for a wide range of possible $\alpha$ and $f_{GW}$ values at a fixed distance = 1 kpc for two different GW detectors : a-LIGO(on the top) and ET(on the bottom). Solid black lines denote contour levels at SNR = 20. } 
\label{fig:SNR}
\end{figure}

\section{Results}
\label{sec:results}
We first consider the signal-to-noise ratio for simulated signals to calculate the detectability and the corresponding errors of the NS parameters associated with possible r-mode detection using the current and future generation GW detectors. To generate the signal model, we consider a canonical NS~\citep{Owen1998} with mass M = 1.4$M_{\odot}$, R = 12.53 km, $\Tilde{J}$ = 0.0163 and I = $10^{38}$ kg-m$^2$ at a distance d = 1 kpc. We consider the r-mode saturation amplitude $\alpha$ in the theoretical expected range of $10^{-6} - 10^{-1}$~\citep{Arras2003,Bondarescu2009} and the CGW frequency in the range of $40-500$ Hz. We don't go up to very high frequency because the universal relations used in our study are not valid for very fast rotating NS~\citep{Doneva_2014,idrisy2015}. We simulate the CGW signals for these wide range of parameter for the NS with a observing period of 2 years which matches the duration of the future LVK observing runs and calculate the expected signal-to-noise ratio (SNR) denoted by $\rho$, using the formula given in Eqn.~\eqref{eq:final_snr} averaged over sky location, polarisation angle and inclination angle. In Fig.~\ref{fig:SNR}, we plot the SNR for both a-LIGO and ET design sensitivity curves and see that for third generation detector ET the SNR is order of magnitude higher, as expected. Considering a minimum SNR of 20 is required for a signal to be detectable, we see that a canonical NS at a distance $1$ kpc with r-mode frequency $\geq 100$ Hz and $\alpha \geq 10^{-4}$ will be detectable with 2 years of observing using the third generation GW detector ET. The SNR estimates shown in this figure also re-scale accordingly with changing distance. \\

\begin{figure}
\centering
\includegraphics[width=1\linewidth]{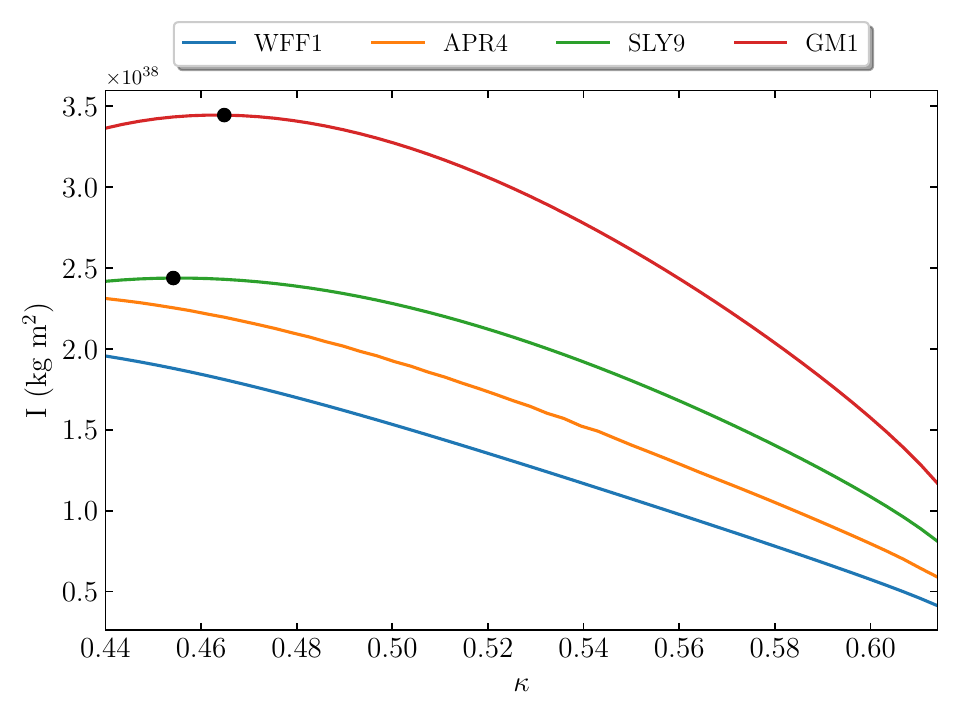}
\caption{Estimated MoI as a function of $\kappa$ for 4 different choice of EOS}
\label{fig:MoI}
\end{figure}

If we want to measure distance of the NS from the detected signal, we need to first estimate the MoI to break their degeneracy as shown in Eqn.~\eqref{eq:moid}. In the case of targeted searches, for detectable signals we will be able to determine the value of r-mode frequency ($f_{GW}$) with a great accuracy from the phase of the signal~\citep{JK1999} and subsequently the value of $\kappa$ using the Eqn.~\ref{eq:freq}. Assuming the sources are slowly rotating NS, we can use the universal relations given in Eqn. ~\ref{eq:univ_lamb} and ~\ref{eq:univ_IloveQ} to estimate the tidal deformability($\Lambda$) and normalised MoI($\Bar{I}$) respectively. To estimate the MoI from the normalised MoI ($\Bar{I}$), we need to assume a true EOS of the NS matter which allows us to calculate the mass of the star from the estimated tidal deformability.  For this analysis, we choose 4 tabulated EOSs : WFF1, APR4, SLY9 and GM1 among which WFF1 and GM1 are the softest and stiffest EOS respectively. All our EOS tables are obtained from either CompOSE ~\citep{Compose,Oertel2016} or an EOS catalogue from~\cite{lalsim} used in LALSuite~\citep{lalsuite}. All the EOSs satisfy the constraints of maximum mass $\geq 2M_{\odot}$, $90\%$ tidal deformability limits from GW170817 (except the very stiff EOS GM1) and the M-R estimates from the NICER measurements~\citep{Abbott_2020,Biswas2022}. For these 4 chosen EOSs, we calculate the MoI using the universal relations and plot them as a function of $\kappa$ in Fig.~\ref{fig:MoI}. For stiff EOSs GM1 and SLY9, we see the unstable branch (starting point is denoted by black dots in the Fig.~\ref{fig:MoI}) determined by ``turning point'' method~\citep{TP} at low values of $\kappa$ which corresponds to high value of compactness that cannot be obtained for stable NSs with these stiff EOSs. \\

\begin{figure}
\centering
\includegraphics[width=1\linewidth]{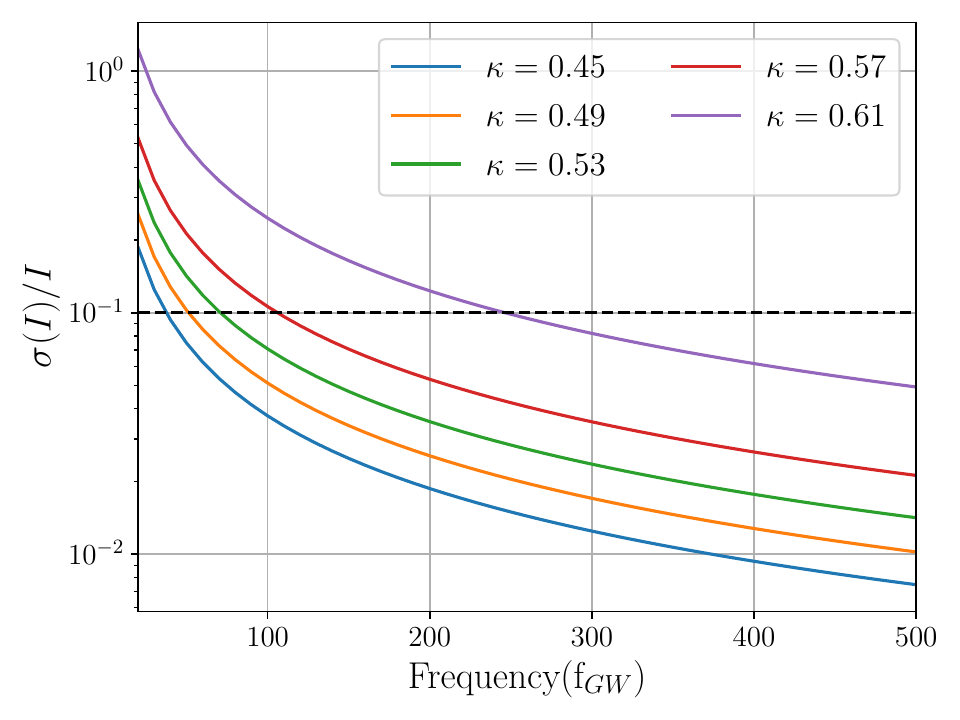}
\caption{Error estimation of the MoI at a fixed value of distance($d$) = 1 kpc with $\alpha = 10^{-3}$ for ET design sensitivity curve for SLY9 EoS. The dashed black lines denote a fixed value of relative error of 10$\%$.}
\label{fig:I_error}
\end{figure}

For the error measurement of the MoI, we assume a fixed pulsar at a distance($d$) = 1 kpc with $\alpha = 10^{-3}$ and consider signals with SNR $> 20$ for the ET design sensitivity so that the signals are detectable. As given in Eqn.~\ref{eq:error_I}, for the error in measurement of $I$  we also require the values of $\kappa$, error in measurement of $f_{rot}$ and choice of EOS. For targeted searches, we generally can measure the rotational frequency($f_{rot}$) of the pulsar from prior EM observations up to a very high precision~\citep{faccuracy_1,faccuracy2}; although we consider here an error of 1$\%$. In Fig.~\ref{fig:I_error}, we plot the error in measuring the parameter $I$ as a function of $f_{GW}$ for different choices of $\kappa$ for the particular EOS, SLY9. We see that the relative error decreases with higher $\kappa$ value but don't see much change in the error estimation with different EOSs(Figures not shown). Also, we observe that for reasonable values of $\kappa$~\citep{Ghosh2023}, we can measure the MoI upto $10\%$ accuracy for CGW frequency $\geq 100$ Hz. We also find that the relative errors will increase for increasing distance and decreasing $\alpha$ (Figures not shown) which is also evident from Eqn.~\eqref{eq:ramp} and Fig.~\ref{fig:SNR} respectively. \\

Although we can estimate the MoI from the frequency measurement using universal relation, to get the distance of the pulsar we need to estimate the characteristic amplitude $h_0$. The differing amplitude of the two polarizations of the gravitational waveform $A_{+}$ and $A_{\times}$ allow us to determine both the binary inclination($\iota$) and the characteristic amplitude $h_0$ from Eqn.~\eqref{eq:incl}. However, both polarizations have nearly identical amplitudes at small inclination angles (difference less than $5\%$ for $\iota \leq 45^{\circ}$) and significantly lower amplitudes at large inclination angles~\citep{Usman_2019}. This leads to the conclusion that the CGW signal is strongest for NSs that are close to face-on ($\iota \sim 0^{\circ}$) or face-away ($\iota \sim 180^{\circ}$) and thus there is an observational bias towards detecting NSs whose orbital angular momentum is well-aligned (or anti-aligned) with the line of sight. In that case, if the amplitudes of the two polarizations are close to equal, we cannot measure strain amplitude or inclination separately. This will lead to a degeneracy between the distance estimation with the inclination angle of the NS which is also present in the case of binary systems~\citep{Nissanke_2010,Schutz_2011}. \\

\begin{figure}
\centering
\includegraphics[width=.5\textwidth]{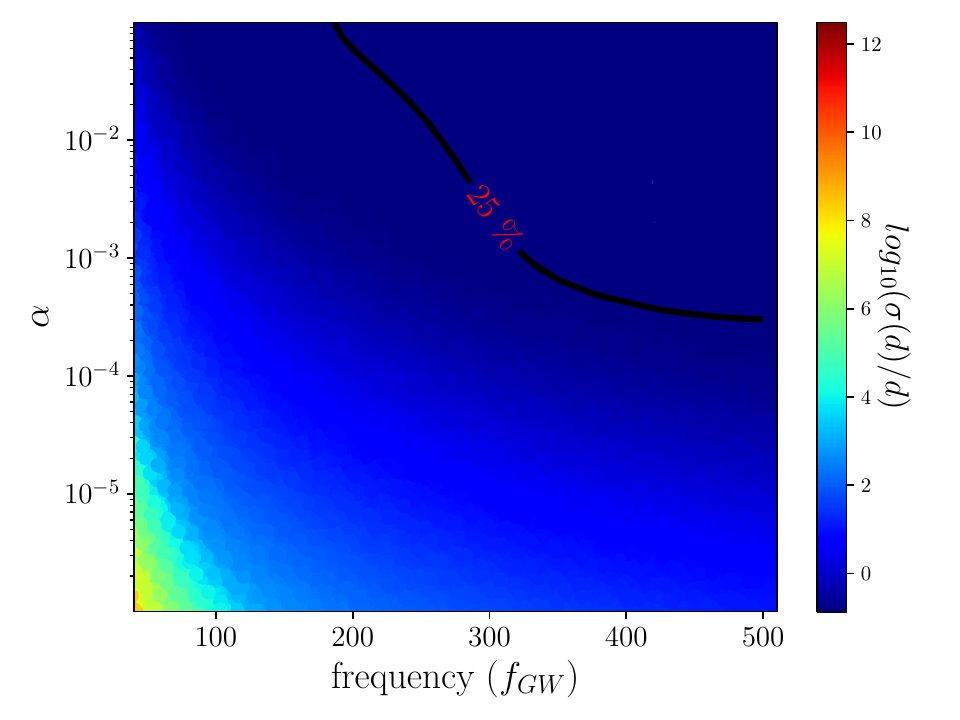}\hfill
\includegraphics[width=.5\textwidth]{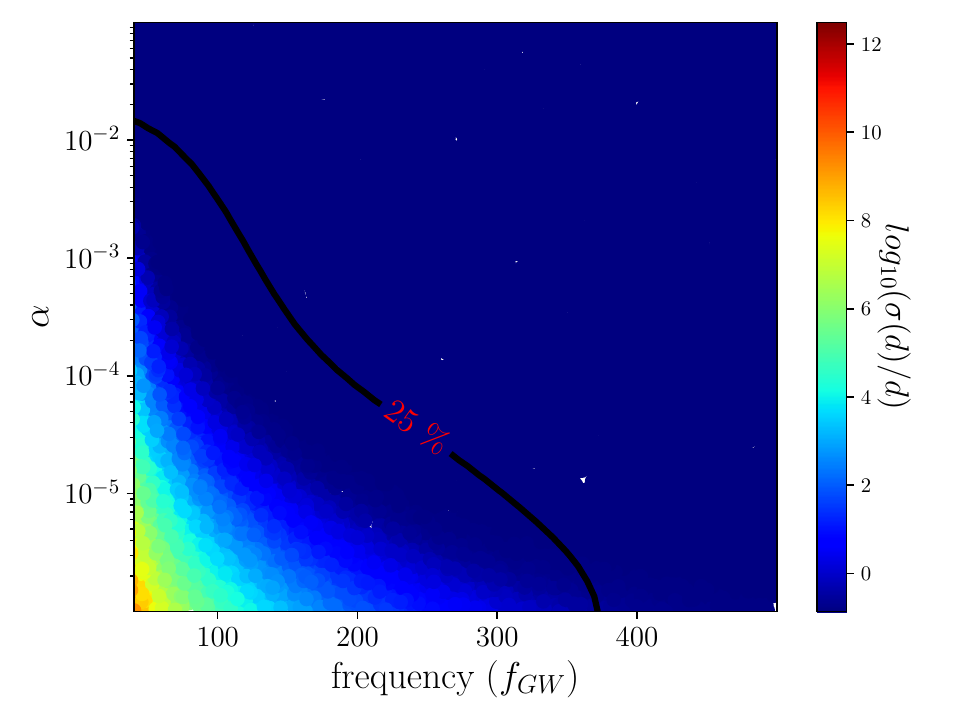}
\caption{Error estimation of $d$ for a wide range of possible $\alpha$ and $f_{GW}$ values at fixed value of $d = 1$ kpc for ET design sensitivity for two different inclination angle : $\iota \sim 0^{\circ}$(on the top) and $\iota \sim 90^{\circ}$ (on the bottom). The solid black  lines denote contours of relative error of $25\%$.} 
\label{fig:d_error}
\end{figure}

To look into the effect of uncertainty of the inclination angle on the distance measurement, in Fig~\ref{fig:d_error} we plot the error in measurement of $d$ as a function of $\alpha$ varied in the range of $10^{-6} - 10^{-1}$ and the CGW frequency to be in the range of $40-500$ Hz for two different inclination : $\iota \sim 0^{\circ}$ and $\iota \sim 90^{\circ}$. A canonical NS at a distance of $1$ kpc with `SLY9' EOS and fixed $\kappa = 0.5$ was considered  for the ET design sensitivity. In the Fig.~\ref{fig:d_error}, we also show contour of $25\%$ relative error in measured distance in black solid lines. For a pulsar at $d = 1$ kpc aligned with the line of sight, for detectable signals we can measure the distance up to an accuracy of $25\%$ for $f_{GW} \geq 300$ Hz and $\alpha \geq 10^{-3}$ but the errors decrease when the inclination angle changes to $\iota \sim 90^{\circ}$. Although SNR decreases with increasing inclination angle changes from $\iota \sim 0^{\circ}$ to $\iota \sim 90^{\circ}$(from Eqn.~\eqref{eq:final_snr_pol}), since the measurement of $h_0$ becomes degenerate with inclination angle at low values, the error in measuring $h_0$ and hence, distance$(d)$ become larger at low inclination angle. These error estimates also scale accordingly with changing distance of the pulsar (Figures not shown). 

\section{Assumptions}
\label{sec:assumptions}
\color{black}
In this section, we revisit some of the assumptions we made for our analysis. Throughout our calculation, we have assumed the entire spin-down is driven by CGW emission via r-modes. This is of course not true for targeted searches, since we already observe them as pulsars and so, only a fraction of the spin-down energy is radiated by CGW emission. A measurement of braking index, $n = \frac{f\Ddot{f}}{\Dot{f}}$ can differentiate between the spin-down mechanisms; for example $n = 3$ indicate dipole emission and $n = 7$ for r-mode emission~\citep{Riles_2023}. Recent analysis of PSR J0537 measures a braking index $n \approx 7$ which indicates that a significant fraction of the spin-down energy is radiated via r-modes~\citep{Anderrson2018}. That is what also make this pulsar most promising candidate for r-mode searches. It may in fact be possible to carry out an analysis assuming a electromagnetic component to the spin-down, as described in~\cite{Lu_2023}. From a successful CGW detection, we will also be able to measure the braking index to great accuracy that can shed light to the various spin-down components~\citep{Sieniawska2021}. \\

While considering the phase of the signal model in~\eqref{eq:phase}, we ignored dependence of the signal on the source’s sky location~\citep{JK1999}. For targeted searches, we have prior knowledge about the sky location of the source and also in general for in CGW observations, sky location is expected to be measured extremely accurately. In practice, we should also consider effect of the cosmological parameters such as redshift($z$) in the detectability of the signals specially when trying to estimate distances. But~\cite{Sieniawska2021} showed that percentage change in the CGW amplitude estimation due to the cosmological corrections are within a few percents and so will not be very important for such detection. To estimate the errors, rather than a full Bayesian analysis, we do a Fisher information matrix formalism which is strictly valid only in the case of high signal-to-noise ratios and also has several other limitations~\citep{vallisneri2008}. Still we adapt this formalism because it is computationally easy to implement and gives a qualitative accurate estimation of the inferred parameters. \\

While considering the signal strain amplitude, we considered a constant value of r-mode  amplitude $\alpha$. If the NS is spinning down via r-mode emission, r-mode amplitude will grow due to the CFS mechanism~\citep{CFS1,CFS2} till it reaches a constant value called saturation amplitude $\alpha_s$ due to dynamical couplings with other modes~\citep{Arras2003}. In our model, we assumed a constant value of $\alpha_s$ similar to~\cite{Owen1998} but different studies indicate more complex behaviour with different limits to the value of $\alpha_s$~\citep{Arras2003,Brink2004,Bondarescu2009}. An upper limit of this saturation amplitude was also obtained from the non-detection of r-modes from PSR J0537 using recent analysis of the O3 data~\citep{LIGO_rmode}. \\

The universal relations used in this paper are also valid under some specific conditions. The universal relations obtained in our recent work~\cite{Ghosh2023} are valid for slow rotation limits only. The higher order effect is of the order of $\left(\frac{f}{f_K}\right)^2$~\citep{idrisy2015} where $f_K$ is the Keplerian frequency($\propto \sqrt{M/R^3}$). The ``I-Love-Q'' relations also do not hold true for rapidly rotating stars~\citep{Doneva_2014} or highly magnetised stars~\citep{Haskell_2013} and become EOS dependent. There are several other factors like the presence of solid crust~\citep{Yuri2001}, stratification~\citep{Yoshida2000,Passamonti2009,Gittins_2023}, magnetic field~\citep{Ho2000,Luciano1,Luciano2,Luciano3,Morsink2002} or superfluidity in the core~\citep{Lindblom2000,Andersson2001} that might affect the r-mode frequency but their effect was found to be negligible for most stars~\citep{idrisy2015}. The most promising target PSR J0537-6910 has rotational frequency = 62 Hz~\citep{Marshall_1998} which determine r-mode frequency in the range of $86 - 99$ Hz~\citep{Ghosh2023}, much lower than its Keplerian frequency. So, for a future detection of r-mode from this particular pulsar, we can use both the universal relations to infer the NS intrinsic properties.  \\

\section{Discussion and Conclusion}
\label{sec:discussion}
In this work, we showed how to measure the various neutron star intrinsic properties from a future successful r-mode detection using the current and third generation GW detectors. We mainly focused on how to break the degeneracy between the MoI and distance measurement from a r-mode detection that was recently pointed out in~\cite{Sieniawska2021}. Using the universal relations obtained in our earlier work~\citep{Ghosh2023}, for targeted pulsars we can measure the compactness and the dimensionless tidal deformability from the measured CGW frequency. Using the ``I-love-Q'' relations, we can then calculate the normalised MoI and knowledge of the true EOS allows to calculate the MoI($I$) of the system. Stiffer EOS predicts higher value of MoI which is expected as stiffer EOS predicts higher radius for a NS. Once the MoI is known, we can easily calculate the distance($d$) from the measured strain amplitude($h_0$). This way, from the measurement of the characteristic strain amplitude($h_0$) and the CGW frequency($f_{GW}$), we can measure the distance of the pulsar using these continuous gravitational observations and break the degeneracy with MoI. But this way of distance measurement still suffers from its degeneracy with inclination angle similar to binary GW observations~\citep{Nissanke_2010, Schutz_2011}. In future, the network of five detectors with the recent addition of KAGRA~\citep{Aso2013} and upcoming LIGO-India~\citep{Saleem_2021} or the triangular configuration of the ET~\citep{ET1} would further increase the network’s sensitivity to constrain both the inclination angle and distance~\citep{Usman_2019}. Electromagnetic observations of the pulsars can also be used to constrain the inclination angle and thus breaking their degeneracy with distance~\citep{Benli_2021}. \\

Although we can calculate the MoI upto a 10\% accuracy for CGW signals from a pulsar at d = 1 kpc with frequency $\geq 100$ Hz; the distance estimation, being dependent on the measurement of $h_0$ has much larger errors and also suffers from its degeneracy with the inclination angle. For pulsars aligned with line of sight, to have below 25\% accuracy in the distance estimation with CGW  frequency $\geq 300$ Hz, we need $\alpha \geq 10^{-3}$. Recent limits on the r-mode saturation amplitudes~\citep{Bondarescu2009,Haskell_2014} suggest that we might need much longer observation periods than considered here with third generation GW detectors to achieve this accuracy. For nearby pulsars, we expect a detection from less observing period with errors in distance measurement comparable to the same from EM observation of pulsars~\citep{Kaplan_2008}. Recently~\cite{Sieniawska_2023} discussed the measurement of distance from CGW observations using  parallax method. Although this parallax method does not require any prior EM knowledge of the pulsar to calculate distance, with this method we can measure distance up to very nearby pulsars (upto few hundreds of parsecs) with sufficient accuracy~\citep{Sieniawska_2023}. Although unlike the case of binary systems, here we need an prior knowledge of the star's rotational frequency from EM observations and the assumption of a true EOS, with the sensitivity of the current searches, this method gives an alternative way to calculate the pulsar distances from CGW observations alone. \\

Another alternate way to use the universal relations is to constrain the NS EOS with the prior knowledge of the distance for the targeted pulsars from EM observations. For the recent r-mode targeted pulsars, PSR J0537-6910~\citep{LIGO_rmode} or the Crab pulsar~\citep{Rajbhandari2021}, we know their distance to good accuracy from the pulsar timing data~\citep{Pietrzy_ski_2019,Kaplan_2008}. Measurement of the MoI of the NS using a prior knowledge of distance and normalised MoI from the CGW frequency using the universal relations as described in Sec.~\ref{sec:results} allows to measure the mass of the NS also (See Fig. 19 for upper limits of the mass estimated from the recent O3 analysis of PSR J0537 by LVK collaboration~\citep{Abbott_2021}). A separate MoI measurement from these r-mode CGW observation along with mass measurements can constrain the dense matter EOS inside NS~\citep{Bejger2003}. It can also be used to check the validity of the ``I-Love-Q'' relation which also carries the signatures of quark matter inside the NS~\citep{yagiYunes2013,Yagi_2017}. This kind of inference also relies on the assumption of saturated r-mode driven spindown and accurate measurement of inclination angle from the CGW detection. \\

Future work could extend to a much more realistic model of r-mode oscillation inside the NS considering both the growth and saturation of r-modes. According to~\cite{Sieniawska2021} a time varying $\alpha$ could also be easily implemented in the energy conservation equations. Recent studies by ~\cite{Gittings_2023b,Gittins_2023} also emphasises on the importance of composition stratification in a mature neutron star for the r-mode oscillations. Also, we should consider the parameter estimation in a full Bayesian framework to give robust conclusions about the inferred NS parameters, as well as to use the prior information from the electromagnetic observations of neutron stars in a consistent way. \\

\section*{Acknowledgements}
 I would like to thank the anonymous referee for the useful comments and suggestions that have helped improve the paper. I would like to thank Prof. Debarati Chatterjee for her encouragement, useful comments and discussions during this work. I would like to thank Bikram Keshari Pradhan and Dhruv Pathak for various discussions during this work. I also acknowledge usage of the IUCAA HPC computing facility, PEGASUS for the numerical calculations.

\section*{Data Availability}
The data underlying this article will be shared on reasonable request to the corresponding author.



\bibliographystyle{mnras}
\bibliography{example} 

\begin{thebibliography}{}
\makeatletter
\relax
\def\mn@urlcharsother{\let\do\@makeother \do\$\do\&\do\#\do\^\do\_\do\%\do\~}
\def\mn@doi{\begingroup\mn@urlcharsother \@ifnextchar [ {\mn@doi@}
  {\mn@doi@[]}}
\def\mn@doi@[#1]#2{\def\@tempa{#1}\ifx\@tempa\@empty \href
  {http://dx.doi.org/#2} {doi:#2}\else \href {http://dx.doi.org/#2} {#1}\fi
  \endgroup}
\def\mn@eprint#1#2{\mn@eprint@#1:#2::\@nil}
\def\mn@eprint@arXiv#1{\href {http://arxiv.org/abs/#1} {{\tt arXiv:#1}}}
\def\mn@eprint@dblp#1{\href {http://dblp.uni-trier.de/rec/bibtex/#1.xml}
  {dblp:#1}}
\def\mn@eprint@#1:#2:#3:#4\@nil{\def\@tempa {#1}\def\@tempb {#2}\def\@tempc
  {#3}\ifx \@tempc \@empty \let \@tempc \@tempb \let \@tempb \@tempa \fi \ifx
  \@tempb \@empty \def\@tempb {arXiv}\fi \@ifundefined
  {mn@eprint@\@tempb}{\@tempb:\@tempc}{\expandafter \expandafter \csname
  mn@eprint@\@tempb\endcsname \expandafter{\@tempc}}}

\bibitem[\protect\citeauthoryear{Aasi et~al.,}{Aasi et~al.}{2015}]{AdvLIGO2015}
Aasi J.,  et~al., 2015, \mn@doi [Classical and Quantum Gravity]
  {10.1088/0264-9381/32/7/074001}, 32, 074001

\bibitem[\protect\citeauthoryear{Abbott et~al.,}{Abbott
  et~al.}{2016a}]{LIGOGW1st}
Abbott R.,  et~al., 2016a, \mn@doi [Phys. Rev. Lett.]
  {10.1103/PhysRevLett.116.061102}, 116, 061102

\bibitem[\protect\citeauthoryear{Abbott et~al.,}{Abbott
  et~al.}{2016b}]{Abbott2016}
Abbott B.~P.,  et~al., 2016b, \mn@doi [Phys. Rev. Lett.]
  {10.1103/PhysRevLett.116.131103}, 116, 131103

\bibitem[\protect\citeauthoryear{Abbott et~al.,}{Abbott et~al.}{2017a}]{CE}
Abbott B.~P.,  et~al., 2017a, \mn@doi [Classical and Quantum Gravity]
  {10.1088/1361-6382/aa51f4}, 34, 044001

\bibitem[\protect\citeauthoryear{Abbott et~al.,}{Abbott
  et~al.}{2017b}]{Abbott2017}
Abbott B.,  et~al., 2017b, \mn@doi [Physical Review Letters]
  {10.1103/physrevlett.119.161101}, 119, 161101

\bibitem[\protect\citeauthoryear{Abbott et~al.,}{Abbott
  et~al.}{2017c}]{MultiGW170817}
Abbott B.,  et~al., 2017c, \mn@doi [The Astrophysical Journal Letters]
  {10.3847/2041-8213/aa91c9}, 848, L12

\bibitem[\protect\citeauthoryear{Abbott et~al.,}{Abbott et~al.}{2018a}]{a-LIGO}
Abbott B.~P.,  et~al., 2018a, Updated Advanced LIGO sensitivity design curve,
  \url{https://dcc.ligo.org/LIGO-T1800044/public}

\bibitem[\protect\citeauthoryear{Abbott et~al.,}{Abbott
  et~al.}{2018b}]{Abbott2018}
Abbott B.,  et~al., 2018b, \mn@doi [Physical Review Letters]
  {10.1103/physrevlett.121.161101}, 121, 161101

\bibitem[\protect\citeauthoryear{Abbott et~al.,}{Abbott
  et~al.}{2019}]{Abbott2019}
Abbott B.,  et~al., 2019, \mn@doi [Phys. Rev. X] {10.1103/PhysRevX.9.011001},
  9, 011001

\bibitem[\protect\citeauthoryear{Abbott et~al.,}{Abbott et~al.}{2020a}]{ALIGO}
Abbott R.,  et~al., 2020a, \mn@doi [Living Reviews in Relativity]
  {10.1007/s41114-020-00026-9}, 23

\bibitem[\protect\citeauthoryear{Abbott et~al.,}{Abbott
  et~al.}{2020b}]{Abbott_2020}
Abbott B.,  et~al., 2020b, \mn@doi [Classical and Quantum Gravity]
  {10.1088/1361-6382/ab5f7c}, 37, 045006

\bibitem[\protect\citeauthoryear{Abbott et~al.,}{Abbott
  et~al.}{2021a}]{GWcatalouge}
Abbott R.,  et~al., 2021a, \mn@doi [Phys. Rev. X] {10.1103/PhysRevX.11.021053},
  11, 021053

\bibitem[\protect\citeauthoryear{Abbott et~al.,}{Abbott
  et~al.}{2021b}]{Abbott_2021}
Abbott R.,  et~al., 2021b, \mn@doi [The Astrophysical Journal Letters]
  {10.3847/2041-8213/ac082e}, 915, L5

\bibitem[\protect\citeauthoryear{Abbott et~al.,}{Abbott
  et~al.}{2021c}]{LIGO_rmode}
Abbott R.,  et~al., 2021c, \mn@doi [The Astrophysical Journal]
  {10.3847/1538-4357/ac0d52}, 922, 71

\bibitem[\protect\citeauthoryear{Acernese et~al.,}{Acernese
  et~al.}{2014}]{AdvVIRGO2014}
Acernese F.,  et~al., 2014, \mn@doi [Classical and Quantum Gravity]
  {10.1088/0264-9381/32/2/024001}, 32, 024001

\bibitem[\protect\citeauthoryear{Akutsu et~al.,}{Akutsu et~al.}{2021}]{Kagra}
Akutsu T.,  et~al., 2021, \mn@doi [Progress of Theoretical and Experimental
  Physics] {10.1093/ptep/ptaa125}, \href
  {https://ui.adsabs.harvard.edu/abs/2021PTEP.2021eA101A} {2021, 05A101}

\bibitem[\protect\citeauthoryear{Alford \& Schwenzer}{Alford \&
  Schwenzer}{2014}]{Alford2014}
Alford M.~G.,  Schwenzer K.,  2014, \mn@doi [The Astrophysical Journal]
  {10.1088/0004-637x/781/1/26}, 781, 26

\bibitem[\protect\citeauthoryear{{Andersson}}{{Andersson}}{1998}]{Andersson1998}
{Andersson} N.,  1998, \mn@doi [The Astrophysical Journal] {10.1086/305919},
  502, 708

\bibitem[\protect\citeauthoryear{Andersson}{Andersson}{2003}]{Andersson2003}
Andersson N.,  2003, \mn@doi [Classical and Quantum Gravity]
  {10.1088/0264-9381/20/7/201}, 20, R105

\bibitem[\protect\citeauthoryear{Andersson \& Comer}{Andersson \&
  Comer}{2001}]{Andersson2001}
Andersson N.,  Comer G.,  2001, \mn@doi [Monthly Notices of the Royal
  Astronomical Society] {10.1046/j.1365-8711.2001.04923.x}, 328, 1129

\bibitem[\protect\citeauthoryear{Andersson \& Gittins}{Andersson \&
  Gittins}{2023}]{Gittings_2023b}
Andersson N.,  Gittins F.,  2023, \mn@doi [The Astrophysical Journal]
  {10.3847/1538-4357/acbc1e}, 945, 139

\bibitem[\protect\citeauthoryear{Andersson, Kokkotas  \& Schutz}{Andersson
  et~al.}{1999a}]{Andersson1999}
Andersson N.,  Kokkotas K.,   Schutz B.~F.,  1999a, \mn@doi [The Astrophysical
  Journal] {10.1086/306625}, 510, 846

\bibitem[\protect\citeauthoryear{Andersson, Kokkotas  \& Stergioulas}{Andersson
  et~al.}{1999b}]{Andersson:1998qs}
Andersson N.,  Kokkotas K.~D.,   Stergioulas N.,  1999b, \mn@doi [Astrophys.
  J.] {10.1086/307082}, 516, 307

\bibitem[\protect\citeauthoryear{Andersson, Antonopoulou, Espinoza, Haskell  \&
  Ho}{Andersson et~al.}{2018}]{Anderrson2018}
Andersson N.,  Antonopoulou D.,  Espinoza C.~M.,  Haskell B.,   Ho W. C.~G.,
  2018, \mn@doi [The Astrophysical Journal] {10.3847/1538-4357/aad6eb}, 864,
  137

\bibitem[\protect\citeauthoryear{{Arras}, {Flanagan}, {Morsink}, {Schenk},
  {Teukolsky}  \& {Wasserman}}{{Arras} et~al.}{2003}]{Arras2003}
{Arras} P.,  {Flanagan} E.~E.,  {Morsink} S.~M.,  {Schenk} A.~K.,  {Teukolsky}
  S.~A.,   {Wasserman} I.,  2003, \mn@doi [The Astrophysical Journal]
  {10.1086/374657}, \href
  {https://ui.adsabs.harvard.edu/abs/2003ApJ...591.1129A} {591, 1129}

\bibitem[\protect\citeauthoryear{Aso, Michimura, Somiya, Ando, Miyakawa,
  Sekiguchi, Tatsumi  \& Yamamoto}{Aso et~al.}{2013}]{Aso2013}
Aso Y.,  Michimura Y.,  Somiya K.,  Ando M.,  Miyakawa O.,  Sekiguchi T.,
  Tatsumi D.,   Yamamoto H.,  2013, \mn@doi [Phys. Rev. D]
  {10.1103/PhysRevD.88.043007}, 88, 043007

\bibitem[\protect\citeauthoryear{{Bejger} \& {Haensel}}{{Bejger} \&
  {Haensel}}{2003}]{Bejger2003}
{Bejger} M.,  {Haensel} P.,  2003, \mn@doi [Astronomy and Astrophysics]
  {10.1051/0004-6361:20030642}, \href
  {https://ui.adsabs.harvard.edu/abs/2003A&A...405..747B} {405, 747}

\bibitem[\protect\citeauthoryear{{Benli, Onur}, {P\'etri, J\'er\^ome}  \&
  {Mitra, Dipanjan}}{{Benli, Onur} et~al.}{2021}]{Benli_2021}
{Benli, Onur} {P\'etri, J\'er\^ome}  {Mitra, Dipanjan} 2021, \mn@doi [A\&A]
  {10.1051/0004-6361/202039853}, 647, A101

\bibitem[\protect\citeauthoryear{Bildsten}{Bildsten}{1998}]{Bildsten:1998ey}
Bildsten L.,  1998, \mn@doi [Astrophys. J. Lett.] {10.1086/311440}, 501, L89

\bibitem[\protect\citeauthoryear{Biswas}{Biswas}{2022}]{Biswas2022}
Biswas B.,  2022, \mn@doi [The Astrophysical Journal]
  {10.3847/1538-4357/ac447b}, 926, 75

\bibitem[\protect\citeauthoryear{{Bondarescu}, {Teukolsky}  \&
  {Wasserman}}{{Bondarescu} et~al.}{2009}]{Bondarescu2009}
{Bondarescu} R.,  {Teukolsky} S.~A.,   {Wasserman} I.,  2009, \mn@doi [Phys.
  Rev. D] {10.1103/PhysRevD.79.104003}, \href
  {https://ui.adsabs.harvard.edu/abs/2009PhRvD..79j4003B} {79, 104003}

\bibitem[\protect\citeauthoryear{Brink, Teukolsky  \& Wasserman}{Brink
  et~al.}{2004}]{Brink2004}
Brink J.,  Teukolsky S.~A.,   Wasserman I.,  2004, \mn@doi [Phys. Rev. D]
  {10.1103/PhysRevD.70.121501}, 70, 121501

\bibitem[\protect\citeauthoryear{Caride, Inta, Owen  \& Rajbhandari}{Caride
  et~al.}{2019}]{Caride2019}
Caride S.,  Inta R.,  Owen B.~J.,   Rajbhandari B.,  2019, \mn@doi [Phys. Rev.
  D] {10.1103/PhysRevD.100.064013}, 100, 064013

\bibitem[\protect\citeauthoryear{Chandrasekhar}{Chandrasekhar}{1970}]{CFS1}
Chandrasekhar S.,  1970, \mn@doi [Phys. Rev. Lett.]
  {10.1103/PhysRevLett.24.611}, 24, 611

\bibitem[\protect\citeauthoryear{Desvignes et~al.,}{Desvignes
  et~al.}{2016}]{faccuracy2}
Desvignes G.,  et~al., 2016, \mn@doi [Monthly Notices of the Royal Astronomical
  Society] {10.1093/mnras/stw483}, 458, 3341

\bibitem[\protect\citeauthoryear{Dietrich, Coughlin, Pang, Bulla, Heinzel,
  Issa, Tews  \& Antier}{Dietrich et~al.}{2020}]{Dietrich2020}
Dietrich T.,  Coughlin M.~W.,  Pang P. T.~H.,  Bulla M.,  Heinzel J.,  Issa L.,
   Tews I.,   Antier S.,  2020, \mn@doi [Science] {10.1126/science.abb4317},
  370, 1450–1453

\bibitem[\protect\citeauthoryear{Doneva, Yazadjiev, Stergioulas  \&
  Kokkotas}{Doneva et~al.}{2013}]{Doneva_2014}
Doneva D.~D.,  Yazadjiev S.~S.,  Stergioulas N.,   Kokkotas K.~D.,  2013,
  \mn@doi [The Astrophysical Journal Letters] {10.1088/2041-8205/781/1/L6},
  781, L6

\bibitem[\protect\citeauthoryear{Evans et~al.,}{Evans et~al.}{2021}]{CE2}
Evans M.,  et~al., 2021, A Horizon Study for Cosmic Explorer: Science,
  Observatories, and Community (\mn@eprint {arXiv} {2109.09882})

\bibitem[\protect\citeauthoryear{Fesik \& Papa}{Fesik \&
  Papa}{2020a}]{Fesik2020a}
Fesik L.,  Papa M.~A.,  2020a, \mn@doi [The Astrophysical Journal]
  {10.3847/1538-4357/ab8193}, 895, 11

\bibitem[\protect\citeauthoryear{Fesik \& Papa}{Fesik \&
  Papa}{2020b}]{Fesik2020b}
Fesik L.,  Papa M.~A.,  2020b, \mn@doi [The Astrophysical Journal]
  {10.3847/1538-4357/aba04e}, 897, 185

\bibitem[\protect\citeauthoryear{Friedman \& Morsink}{Friedman \&
  Morsink}{1998}]{Friedman:1997uh}
Friedman J.~L.,  Morsink S.~M.,  1998, \mn@doi [Astrophys. J.]
  {10.1086/305920}, 502, 714

\bibitem[\protect\citeauthoryear{{Friedman} \& {Schutz}}{{Friedman} \&
  {Schutz}}{1978}]{CFS2}
{Friedman} J.~L.,  {Schutz} B.~F.,  1978, \mn@doi [The Astrophysical Journal]
  {10.1086/156143}, 222, 281

\bibitem[\protect\citeauthoryear{{Friedman}, {Ipser}  \& {Sorkin}}{{Friedman}
  et~al.}{1988}]{TP}
{Friedman} J.~L.,  {Ipser} J.~R.,   {Sorkin} R.~D.,  1988, \mn@doi [The
  Astrophysical Journal] {10.1086/166043}, \href
  {https://ui.adsabs.harvard.edu/abs/1988ApJ...325..722F} {325, 722}

\bibitem[\protect\citeauthoryear{Ghosh, Pradhan, Chatterjee  \&
  Schaffner-Bielich}{Ghosh et~al.}{2022a}]{Ghosh_2022}
Ghosh S.,  Pradhan B.~K.,  Chatterjee D.,   Schaffner-Bielich J.,  2022a,
  \mn@doi [Front. Astron. Space Sci.] {10.3389/fspas.2022.864294}, 9, 864294

\bibitem[\protect\citeauthoryear{Ghosh, Chatterjee  \& Schaffner-Bielich}{Ghosh
  et~al.}{2022b}]{Ghosh2022}
Ghosh S.,  Chatterjee D.,   Schaffner-Bielich J.,  2022b, \mn@doi [Eur. Phys.
  J. A] {10.1140/epja/s10050-022-00679-w}, 58, 37

\bibitem[\protect\citeauthoryear{Ghosh, Pathak  \& Chatterjee}{Ghosh
  et~al.}{2023}]{Ghosh2023}
Ghosh S.,  Pathak D.,   Chatterjee D.,  2023, \mn@doi [The Astrophysical
  Journal] {10.3847/1538-4357/acb0d3}, 944, 53

\bibitem[\protect\citeauthoryear{Gittins \& Andersson}{Gittins \&
  Andersson}{2023}]{Gittins_2023}
Gittins F.,  Andersson N.,  2023, \mn@doi [Monthly Notices of the Royal
  Astronomical Society] {10.1093/mnras/stad672}, 521, 3043

\bibitem[\protect\citeauthoryear{Haskell, Ciolfi, Pannarale  \&
  Rezzolla}{Haskell et~al.}{2013}]{Haskell_2013}
Haskell B.,  Ciolfi R.,  Pannarale F.,   Rezzolla L.,  2013, \mn@doi [Monthly
  Notices of the Royal Astronomical Society: Letters] {10.1093/mnrasl/slt161},
  438, L71

\bibitem[\protect\citeauthoryear{Haskell, Glampedakis  \& Andersson}{Haskell
  et~al.}{2014}]{Haskell_2014}
Haskell B.,  Glampedakis K.,   Andersson N.,  2014, \mn@doi [Monthly Notices of
  the Royal Astronomical Society] {10.1093/mnras/stu535}, 441, 1662

\bibitem[\protect\citeauthoryear{Hild et~al.,}{Hild et~al.}{2011}]{ET2}
Hild S.,  et~al., 2011, \mn@doi [Classical and Quantum Gravity]
  {10.1088/0264-9381/28/9/094013}, 28, 094013

\bibitem[\protect\citeauthoryear{Ho \& Lai}{Ho \& Lai}{2000}]{Ho2000}
Ho W. C.~G.,  Lai D.,  2000, \mn@doi [The Astrophysical Journal]
  {10.1086/317085}, 543, 386

\bibitem[\protect\citeauthoryear{Ho, Andersson  \& Haskell}{Ho
  et~al.}{2011}]{Wynn2011}
Ho W. C.~G.,  Andersson N.,   Haskell B.,  2011, \mn@doi [Phys. Rev. Lett.]
  {10.1103/PhysRevLett.107.101101}, 107, 101101

\bibitem[\protect\citeauthoryear{Idrisy, Owen  \& Jones}{Idrisy
  et~al.}{2015}]{idrisy2015}
Idrisy A.,  Owen B.~J.,   Jones D.~I.,  2015, \mn@doi [Physical Review D]
  {10.1103/physrevd.91.024001}, 91

\bibitem[\protect\citeauthoryear{Jaranowski \& Kr\'olak}{Jaranowski \&
  Kr\'olak}{1999}]{JK1999}
Jaranowski P.,  Kr\'olak A.,  1999, \mn@doi [Phys. Rev. D]
  {10.1103/PhysRevD.59.063003}, 59, 063003

\bibitem[\protect\citeauthoryear{Jaranowski, Kr\'olak  \& Schutz}{Jaranowski
  et~al.}{1998}]{JK1998}
Jaranowski P.,  Kr\'olak A.,   Schutz B.~F.,  1998, \mn@doi [Phys. Rev. D]
  {10.1103/PhysRevD.58.063001}, 58, 063001

\bibitem[\protect\citeauthoryear{Kaplan, Chatterjee, Gaensler  \&
  Anderson}{Kaplan et~al.}{2008}]{Kaplan_2008}
Kaplan D.~L.,  Chatterjee S.,  Gaensler B.~M.,   Anderson J.,  2008, \mn@doi
  [The Astrophysical Journal] {10.1086/529026}, 677, 1201

\bibitem[\protect\citeauthoryear{{LIGO Scientific Collaboration}}{{LIGO
  Scientific Collaboration}}{2018}]{lalsuite}
{LIGO Scientific Collaboration} 2018, {LIGO} {A}lgorithm {L}ibrary -
  {LALS}uite, free software (GPL), \mn@doi{10.7935/GT1W-FZ16}

\bibitem[\protect\citeauthoryear{Levin \& Ushomirsky}{Levin \&
  Ushomirsky}{2001}]{Yuri2001}
Levin Y.,  Ushomirsky G.,  2001, \mn@doi [Monthly Notices of the Royal
  Astronomical Society] {10.1046/j.1365-8711.2001.04323.x}, 324, 917

\bibitem[\protect\citeauthoryear{Lindblom \& Mendell}{Lindblom \&
  Mendell}{2000}]{Lindblom2000}
Lindblom L.,  Mendell G.,  2000, \mn@doi [Phys. Rev. D]
  {10.1103/PhysRevD.61.104003}, 61, 104003

\bibitem[\protect\citeauthoryear{Lindblom, Owen  \& Morsink}{Lindblom
  et~al.}{1998}]{Lindblom1998}
Lindblom L.,  Owen B.~J.,   Morsink S.~M.,  1998, \mn@doi [Phys. Rev. Lett.]
  {10.1103/PhysRevLett.80.4843}, 80, 4843

\bibitem[\protect\citeauthoryear{Lindblom, Mendell  \& Owen}{Lindblom
  et~al.}{1999}]{Lindblom:1999yk}
Lindblom L.,  Mendell G.,   Owen B.~J.,  1999, \mn@doi [Phys. Rev. D]
  {10.1103/PhysRevD.60.064006}, 60, 064006

\bibitem[\protect\citeauthoryear{Lockitch, Andersson  \& Friedman}{Lockitch
  et~al.}{2000}]{Lockitch2001}
Lockitch K.~H.,  Andersson N.,   Friedman J.~L.,  2000, \mn@doi [Phys. Rev. D]
  {10.1103/PhysRevD.63.024019}, 63, 024019

\bibitem[\protect\citeauthoryear{Lockitch, Friedman  \& Andersson}{Lockitch
  et~al.}{2003}]{Lockitch2003}
Lockitch K.~H.,  Friedman J.~L.,   Andersson N.,  2003, \mn@doi [Phys. Rev. D]
  {10.1103/PhysRevD.68.124010}, 68, 124010

\bibitem[\protect\citeauthoryear{Lu, Wette, Scott  \& Melatos}{Lu
  et~al.}{2023}]{Lu_2023}
Lu N.,  Wette K.,  Scott S.~M.,   Melatos A.,  2023, \mn@doi [Monthly Notices
  of the Royal Astronomical Society] {10.1093/mnras/stad390}, 521, 2103

\bibitem[\protect\citeauthoryear{{Manchester} et~al.,}{{Manchester}
  et~al.}{2013}]{faccuracy_1}
{Manchester} R.~N.,  et~al., 2013, \mn@doi [Publications of the Astron. Soc. of
  Australia] {10.1017/pasa.2012.017}, \href
  {https://ui.adsabs.harvard.edu/abs/2013PASA...30...17M} {30, e017}

\bibitem[\protect\citeauthoryear{Marshall, Gotthelf, Zhang, Middleditch  \&
  Wang}{Marshall et~al.}{1998}]{Marshall_1998}
Marshall F.~E.,  Gotthelf E.~V.,  Zhang W.,  Middleditch J.,   Wang Q.~D.,
  1998, \mn@doi [The Astrophysical Journal] {10.1086/311381}, 499, L179

\bibitem[\protect\citeauthoryear{Moore, Cole  \& Berry}{Moore
  et~al.}{2014}]{Moore2015}
Moore C.~J.,  Cole R.~H.,   Berry C. P.~L.,  2014, \mn@doi [Classical and
  Quantum Gravity] {10.1088/0264-9381/32/1/015014}, 32, 015014

\bibitem[\protect\citeauthoryear{Morsink \& Rezania}{Morsink \&
  Rezania}{2002}]{Morsink2002}
Morsink S.~M.,  Rezania V.,  2002, \mn@doi [The Astrophysical Journal]
  {10.1086/341190}, 574, 908

\bibitem[\protect\citeauthoryear{Nissanke, Holz, Hughes, Dalal  \&
  Sievers}{Nissanke et~al.}{2010}]{Nissanke_2010}
Nissanke S.,  Holz D.~E.,  Hughes S.~A.,  Dalal N.,   Sievers J.~L.,  2010,
  \mn@doi [The Astrophysical Journal] {10.1088/0004-637X/725/1/496}, 725, 496

\bibitem[\protect\citeauthoryear{Oertel, Hempel, Klaehn  \& Typel}{Oertel
  et~al.}{2017a}]{Compose}
Oertel M.,  Hempel M.,  Klaehn T.,   Typel S.,  2017a,
  \url{(https://compose.obspm.fr/)}

\bibitem[\protect\citeauthoryear{Oertel, Hempel, Klaehn  \& Typel}{Oertel
  et~al.}{2017b}]{Oertel2016}
Oertel M.,  Hempel M.,  Klaehn T.,   Typel S.,  2017b, \mn@doi [Rev. Mod.
  Phys.] {10.1103/RevModPhys.89.015007}, 89, 015007

\bibitem[\protect\citeauthoryear{Owen}{Owen}{2010}]{Owen2010}
Owen B.~J.,  2010, \mn@doi [Phys. Rev. D] {10.1103/PhysRevD.82.104002}, 82,
  104002

\bibitem[\protect\citeauthoryear{Owen, Lindblom, Cutler, Schutz, Vecchio  \&
  Andersson}{Owen et~al.}{1998}]{Owen1998}
Owen B.~J.,  Lindblom L.,  Cutler C.,  Schutz B.~F.,  Vecchio A.,   Andersson
  N.,  1998, \mn@doi [Phys. Rev. D] {10.1103/PhysRevD.58.084020}, 58, 084020

\bibitem[\protect\citeauthoryear{Pang, Tews, Coughlin, Bulla, Broeck  \&
  Dietrich}{Pang et~al.}{2021}]{Pang_2021}
Pang P. T.~H.,  Tews I.,  Coughlin M.~W.,  Bulla M.,  Broeck C. V.~D.,
  Dietrich T.,  2021, \mn@doi [The Astrophysical Journal]
  {10.3847/1538-4357/ac19ab}, 922, 14

\bibitem[\protect\citeauthoryear{Papaloizou \& Pringle}{Papaloizou \&
  Pringle}{1978}]{rmode1977}
Papaloizou J.,  Pringle J.~E.,  1978, \mn@doi [Monthly Notices of the Royal
  Astronomical Society] {10.1093/mnras/182.3.423}, 182, 423

\bibitem[\protect\citeauthoryear{Passamonti, Haskell, Andersson, Jones  \&
  Hawke}{Passamonti et~al.}{2009}]{Passamonti2009}
Passamonti A.,  Haskell B.,  Andersson N.,  Jones D.~I.,   Hawke I.,  2009,
  \mn@doi [Monthly Notices of the Royal Astronomical Society]
  {10.1111/j.1365-2966.2009.14408.x}, 394, 730

\bibitem[\protect\citeauthoryear{Pietrzy{\'{n}}ski et~al.,}{Pietrzy{\'{n}}ski
  et~al.}{2019}]{Pietrzy_ski_2019}
Pietrzy{\'{n}}ski G.,  et~al., 2019, \mn@doi [Nature]
  {10.1038/s41586-019-0999-4}, 567, 200

\bibitem[\protect\citeauthoryear{Prix}{Prix}{2007}]{Prix2011}
Prix R.,  2007, \mn@doi [Phys. Rev. D] {10.1103/PhysRevD.75.023004}, 75, 023004

\bibitem[\protect\citeauthoryear{{Punturo} et~al.,}{{Punturo}
  et~al.}{2010}]{ET1}
{Punturo} M.,  et~al., 2010, \mn@doi [Classical and Quantum Gravity]
  {10.1088/0264-9381/27/8/084007}, \href
  {https://ui.adsabs.harvard.edu/abs/2010CQGra..27h4007P} {27, 084007}

\bibitem[\protect\citeauthoryear{Rajbhandari, Owen, Caride  \&
  Inta}{Rajbhandari et~al.}{2021}]{Rajbhandari2021}
Rajbhandari B.,  Owen B.~J.,  Caride S.,   Inta R.,  2021, \mn@doi [Physical
  Review D] {10.1103/physrevd.104.122008}, 104

\bibitem[\protect\citeauthoryear{{Rezzolla}, {Lamb}  \& {Shapiro}}{{Rezzolla}
  et~al.}{2000}]{Luciano1}
{Rezzolla} L.,  {Lamb} F.~K.,   {Shapiro} S.~L.,  2000, \mn@doi [Astrophysical
  Journal, Letters] {10.1086/312539}, \href
  {https://ui.adsabs.harvard.edu/abs/2000ApJ...531L.139R} {531, L139}

\bibitem[\protect\citeauthoryear{{Rezzolla}, {Lamb}, {Markovi{\'c}}  \&
  {Shapiro}}{{Rezzolla} et~al.}{2001a}]{Luciano2}
{Rezzolla} L.,  {Lamb} F.~K.,  {Markovi{\'c}} D.,   {Shapiro} S.~L.,  2001a,
  \mn@doi [Phys. Rev. D] {10.1103/PhysRevD.64.104013}, \href
  {https://ui.adsabs.harvard.edu/abs/2001PhRvD..64j4013R} {64, 104013}

\bibitem[\protect\citeauthoryear{{Rezzolla}, {Lamb}, {Markovi{\'c}}  \&
  {Shapiro}}{{Rezzolla} et~al.}{2001b}]{Luciano3}
{Rezzolla} L.,  {Lamb} F.~K.,  {Markovi{\'c}} D.,   {Shapiro} S.~L.,  2001b,
  \mn@doi [Phys. Rev. D] {10.1103/PhysRevD.64.104014}, \href
  {https://ui.adsabs.harvard.edu/abs/2001PhRvD..64j4014R} {64, 104014}

\bibitem[\protect\citeauthoryear{Riles}{Riles}{2023}]{Riles_2023}
Riles K.,  2023, \mn@doi [Living Reviews in Relativity]
  {10.1007/s41114-023-00044-3}, 26, 3

\bibitem[\protect\citeauthoryear{Saleem et~al.,}{Saleem
  et~al.}{2021}]{Saleem_2021}
Saleem M.,  et~al., 2021, \mn@doi [Classical and Quantum Gravity]
  {10.1088/1361-6382/ac3b99}, 39, 025004

\bibitem[\protect\citeauthoryear{Schutz}{Schutz}{1986}]{Schutz1986}
Schutz B.~F.,  1986, \mn@doi [Nature] {10.1038/323310a0}, 323, 310

\bibitem[\protect\citeauthoryear{Schutz}{Schutz}{2011}]{Schutz_2011}
Schutz B.~F.,  2011, \mn@doi [Classical and Quantum Gravity]
  {10.1088/0264-9381/28/12/125023}, 28, 125023

\bibitem[\protect\citeauthoryear{Sieniawska \& Jones}{Sieniawska \&
  Jones}{2021}]{Sieniawska2021}
Sieniawska M.,  Jones D.~I.,  2021, \mn@doi [Monthly Notices of the Royal
  Astronomical Society] {10.1093/mnras/stab3315}, 509, 5179

\bibitem[\protect\citeauthoryear{Sieniawska, Jones  \& Miller}{Sieniawska
  et~al.}{2023}]{Sieniawska_2023}
Sieniawska M.,  Jones D.~I.,   Miller A.~L.,  2023, \mn@doi [Monthly Notices of
  the Royal Astronomical Society] {10.1093/mnras/stad624}

\bibitem[\protect\citeauthoryear{Traversi, Char  \& Pagliara}{Traversi
  et~al.}{2020}]{Traversi2020}
Traversi S.,  Char P.,   Pagliara G.,  2020, \mn@doi [The Astrophysical
  Journal] {10.3847/1538-4357/ab99c1}, 897, 165

\bibitem[\protect\citeauthoryear{Usman, Mills  \& Fairhurst}{Usman
  et~al.}{2019}]{Usman_2019}
Usman S.~A.,  Mills J.~C.,   Fairhurst S.,  2019, \mn@doi [The Astrophysical
  Journal] {10.3847/1538-4357/ab0b3e}, 877, 82

\bibitem[\protect\citeauthoryear{Vallisneri}{Vallisneri}{2008}]{vallisneri2008}
Vallisneri M.,  2008, \mn@doi [Phys. Rev. D] {10.1103/PhysRevD.77.042001}, 77,
  042001

\bibitem[\protect\citeauthoryear{Yagi \& Yunes}{Yagi \&
  Yunes}{2013a}]{yagiYunes2013}
Yagi K.,  Yunes N.,  2013a, \mn@doi [Phys. Rev. D]
  {10.1103/PhysRevD.88.023009}, 88, 023009

\bibitem[\protect\citeauthoryear{Yagi \& Yunes}{Yagi \&
  Yunes}{2013b}]{Yagi2013}
Yagi K.,  Yunes N.,  2013b, \mn@doi [Science] {10.1126/science.1236462}, 341,
  365

\bibitem[\protect\citeauthoryear{Yagi \& Yunes}{Yagi \&
  Yunes}{2017}]{Yagi_2017}
Yagi K.,  Yunes N.,  2017, \mn@doi [Physics Reports]
  {10.1016/j.physrep.2017.03.002}, 681, 1

\bibitem[\protect\citeauthoryear{Yoshida \& Lee}{Yoshida \&
  Lee}{2000}]{Yoshida2000}
Yoshida S.,  Lee U.,  2000, \mn@doi [The Astrophysical Journal Supplement
  Series] {10.1086/313410}, 129, 353

\bibitem[\protect\citeauthoryear{Özel \& Freire}{Özel \&
  Freire}{2016}]{lalsim}
Özel F.,  Freire P.,  2016, \mn@doi [Annual Review of Astronomy and
  Astrophysics] {10.1146/annurev-astro-081915-023322}, 54, 401

\makeatother
\end{thebibliography}







\bsp	
\label{lastpage}
\end{document}